# A study on the kinetic arrest of magnetic phases in nanostructured $Nd_{0.6}Sr_{0.4}MnO_3$ thin films

**R S MRINALENI [1, 2], *E P AMALADASS [1, 2], A. T. SATHYANARAYANA [1, 2], JEGADEESAN P [1], S AMIRTHAPANDIAN [1,2], AND AWADHESH MANI [1,2]**

[1]Material Science Group, Indira Gandhi Centre for Atomic Research, Kalpakkam, 603102

[2]Homi Bhabha National Institute, Kalpakkam, 603102

**Corresponding author: edward@igcar.gov.in*

## Abstract

The $Nd_{0.6}Sr_{0.4}MnO_3$ manganite system exhibits a phase transition from paramagnetic insulating (PMI) to ferromagnetic metallic (FMM) state around its Curie temperature $T_C = 270$ K (bulk). The morphology-driven changes in the kinetically arrested magnetic phases in NSMO thin films with granular and a crossed-nano-rod-type morphology are studied. At low temperatures, the manganite thin films possess a magnetic glassy state arising from the coexistence of the high-temperature PMI phase and the low-temperature FMM phase. The extent of kinetic arrest and its relaxation was studied using the "cooling and heating in unequal field (CHUF)" protocol in magnetic and magneto-transport investigations. The sample with rod morphology showed a large extent of phase coexistence compared to the granular sample. Further, with a field-cooling protocol, time-evolution studies were carried out to understand the relaxation of arrested magnetic phases across these morphologically distinct thin films. The results on the devitrification of the arrested magnetic state are interpreted from the point of view of homogeneous and heterogeneous nucleation of the FM phase in the PM matrix with respect to temperature.

## INTRODUCTION

The manganese oxides commonly referred to as manganites of the general form $RE_{1-x}A_xMnO_3$ are one of the extensively studied strongly correlated electron systems in condensed matter physics[1,2]. Apart from being popular candidates for oxide-based vertically aligned nanostructures (VANs), spintronic devices such as magnetic-tunnel junctions (MTJs), MRAM devices, and FM electrodes, in recent times there has been a surge of interest in these systems due to their interesting applications in the field of magnetic neuromorphic computing, and oxide-based heterostructures hosting non-trivial spin-textures[3–9]. The magnetoresistive effects such as colossal magneto-resistance (CMR), anisotropic magnetoresistance (AMR), low-field magnetoresistance (LFMR), *etc*. have been the highlighting features of manganites[10,11]. Phase coexistence is believed to be the important reason for the observation of the CMR effect[1]. The manganites possess a rich magnetic phase diagram that can be tuned by varying concentrations of the di-valent A- site ion which results in exotic magnetic phases such as paramagnetic (PM), ferromagnetic (FM), antiferromagnetic (AFM) phase and the charge-ordered phase[12]. The $Nd_{0.6}Sr_{0.4}MnO_3$ (NSMO) is a relatively less explored system from the manganite family and exhibits a phase transition from PM insulator to FM metal (IMT) around the Curie temperature $T_C \approx 270$ K (bulk)[13]. The manganites exhibit disorder-induced first-order phase transition[14]. The intrinsic disorder due to stoichiometric doping present in the system is responsible for interesting magnetic phases. However, any disorder in the nano/microscopic range will have its effect on the phase coexistence[14]. In our previous work, we prepared thin films with distinct surface morphologies – granular and nano-rod type[15]. We found that the changes in morphology, influence the magnetic properties of the thin films. Further, the granular thin film showed a non-monotonic behavior at low temperatures in AMR measurements and we attributed it to the presence of kinetically arrested magnetic phases[16]. The main interest of this work is to investigate the magnetic glassy phase across these thin films.

Similar to structural glass, at low temperatures, manganites exhibit a metastable magnetic state termed the magnetic glassy state[17]. This state arises due to the kinetic arrest of the high-temperature magnetic phase along with the existing low-temperature magnetic phase in the system. In this metastable state, the dynamics of the system are frozen (arrested) similar to a structural glass state and it has a large relaxation time which increases with the decrease in temperature below the glass transition temperature '$T_g$'[18]. This has been observed in many magnetic systems where the first-order transition is inhibited and the system cannot attain the equilibrium state[19–24]. As a result, regions of glass-like arrested phase exist up to the lowest temperatures. Only devitrification restores the kinetics which is usually done by heating the sample or with the application of an external field[18]. Other than the kinetically arrested state, another metastable state that can be present in systems undergoing first-order phase transition is the supercooled state[18]. Unlike the kinetically arrested state, the supercooled state is a metastable state with a free-energy barrier inhibiting its conversion to the equilibrium phase (note that the kinetically arrested state does not have any barrier in free energy) [18]. Though the dynamics slow down as the sample is cooled below its supercooling limit (T*), upon lowering the free-energy barrier, this non-equilibrium metastable state can get converted to the equilibrium state[18]. For instance, as the free energy barrier decreases with the reduction in temperature, the metastable supercooled PM state can relax to the equilibrium FM state in our case. The study of kinetically arrested magnetic phases has been of interest in the manganite systems not only from an application point of view but also to expand a basic understanding of glass formation[18]. Previous studies on kinetic arrest have been mainly performed on half-metallic/ $x$ = 0.5 cases in manganites, mostly in bulk samples[21,22,25]. One such work is the substitution of Mn with Al in the Pr-Ca-Mn-O system which has given rise to the special 'cooling and heating in unequal field (CHUF)' protocol[26]. The CHUF protocol is widely used in literature to investigate the phase coexistence and kinetically arrested magnetic phases in manganites. Studies on the glassy FM (AFM) phase in systems with equilibrium AFM (FM) phases have been conducted in the past[19,27]. In our case, the high-temperature phase is PM nature and we aim to investigate its coexistence with the equilibrium ferromagnetic phase in the metastable forms (kinetically arrested and supercooled) at low temperatures.

The formation and evolution of the coexisting magnetic phases due to kinetic arrest depends on many factors in bulk and nanoparticles, but in the case of thin films, the strain in the thin film plays a major role. In previous work, strain is used to populate the surface of manganite thin films with metallic and insulating regions[28]. Further, another study on $La_{5/8-y}Pr_yCa_{3/8}MnO_3$ revealed that the strain developed during the film growth and relaxation process might affect the length scale of phase separation in thin films[29]. In a work by Gayathri *et. al.*, the CHUF protocol is used to examine the presence of FM phases at the interface of the PCMO/PSMO system[30]. These studies on phase coexistence are exceptionally useful in bringing out exotic magnetic properties at manganite interfaces and heterostructures such as the phase separation-driven enhancement in magnetoresistance[30,31]. Having performed extensive magnetotransport investigations in our previous works, we wondered whether the morphology (strain) of these thin films could have an impact on the phase-coexistence and kinetic arrest of magnetic phases. In this work, we have attempted to study the kinetic arrest and devitrification of the glassy PM phases in NSMO thin films. Subsequently, we have performed magnetic and magneto-transport investigations using the CHUF protocol and time-evolution studies on these NSMO thin films.

## EXPERIMENTAL METHODS

A commercial NSMO pellet was used as the target and single crystals of $SrTiO_3$ with (100) orientation were used as substrates. Thin films of NSMO were synthesized using the pulsed laser deposition technique (PLD) with a KrF excimer laser ($\lambda$=248 nm) operating at a fluence of 1.5 J/cm$^2$ with a repetition rate of 3 Hz. The oxygen partial pressure was fixed at 0.36 mbar and the substrate temperature was maintained at 750 $^o$C during deposition. In-situ annealing was carried out after deposition inside the PLD chamber for 2h with an oxygen background pressure of 500 mbar. Further, an ex-situ annealing step was adopted and the thin films were annealed in a tube furnace at 950 $^o$C for 2h in flowing oxygen at a rate of 20 sccm. The 100 nm thick films possessed two types of morphology though prepared under the same experimental conditions. The reason for such nanostructuring was found to be due to the altered growth modes occurring at the substrate surface[15]. The STO substrate possesses a miscut and this leads to steps and terraces at the substrate surface with mixed chemical terminations of SrO and $TiO_2$. When there is partial or incomplete removal of SrO termination, it leads to an island-type growth resulting in a thin film with granular morphology. However, a deliberate $TiO_2$ termination leads to a layer-by-layer growth mode with a resulting nano-rod type morphology. Further, structural and microstructure analysis was performed and it was found that the granular thin film is polycrystalline whereas the film with rod morphology is highly oriented along (00l). A detailed description of changes in growth mode and structural characterization can be found in our previous work[15]. Henceforth, the NSMO thin film with granular morphology will be referred to as NS-G, and the rod-morphology film as NS-R.

The surface morphology of the thin films was determined using scanning electron microscopy (SEM) with crossbeam 340 (from Carl Zeiss). Resistivity and magnetotransport measurements were performed in a magnetoresistance setup by Cryogenics, UK. Magnetization measurements have been performed using a SQUID-based vibrating sample magnetometer system from Quantum Design (EverCool). During the magnetoresistance and magnetization measurements, the sample was mounted such that the magnetic field was parallel to the sample surface.

## RESULTS AND DISCUSSION

Figure 1 a), and c) shows the SEM images of the NSMO thin films chosen in our study. The NSMO thin film in Figure 1 a) exhibits a granular morphology with faceted grains with an average grain size of 34 nm. Figure 1 c) shows the morphology of NSMO thin film where rod-type features are embedded in a well-connected matrix of NSMO. To confirm the presence of phase co-existence up to the lowest temperatures across these samples, we have carried out resistivity and magnetization measurements.

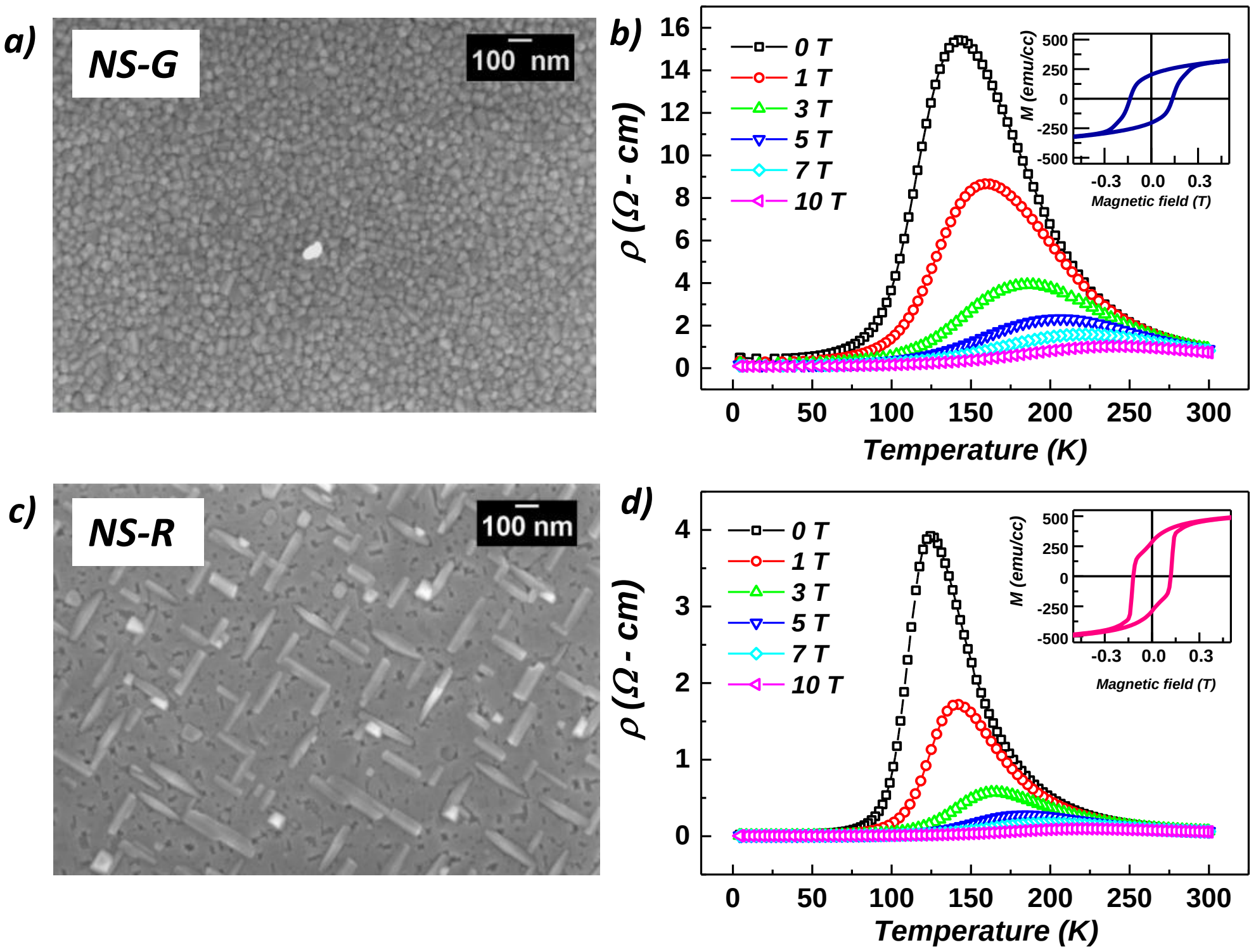

Figure 1 : a) SEM images of NSMO thin films with granular morphology (NS-G) and c) rod-type morphology (NS-R). b), d) Temperature-dependent resistivity of NS-G and NS-R upon cooling the sample in different magnetic fields, respectively. Inset b), d): Magnetization data at 4 K for NS-G and NS-R, respectively.

Temperature-dependent resistivity measurements revealed that an IMT is observed in both the thin films. As shown in Figure 1 b) and d) the granular thin film shows an IMT at 145 K and a sample with rod morphology at 127 K. This IMT temperature is lower than the previously reported value by us[15], as a portion of the thin film was taken to carry out transport measurements. Further on performing resistivity measurements in cooling and warming cycles a thermal hysteresis ($\Delta T \approx 0.5$ K) is observed in both the samples in Figure 2 a), and d). Additionally, the presence of ‘training effect' has also been observed in the MR plots at 4 K for granular and rod-morphology films from Figure 2 b), and d). These observations confirm that the transition is first order in nature and there is phase co-existence up to low temperatures. Further, as the samples are cooled with different magnetic fields, the resistivity measurement as shown in Figure 1 b), d) reveals that the IMT shifts towards higher temperatures with the magnetic field. Such an event is possible only if the FM and PM phases co-exist. This is because the application of magnetic field stabilizes the FM domain formation at higher temperatures and aligns the magnetic moments of the randomly oriented PM clusters leading to the reduction in resistivity and shift the transition point to a higher temperature[2]. Furthermore, due to the presence of increased pinning centres in the granular thin film, NS-G exhibits a higher coercive- field of 1363 Oe than NS-R with a coercive field of 1186 Oe at 4 K (see inset: Figure 1 b), d)). These differences across the samples are very important since defects such as grain -boundaries (GBs) play a vital role in tuning the co-existing phases in the system.

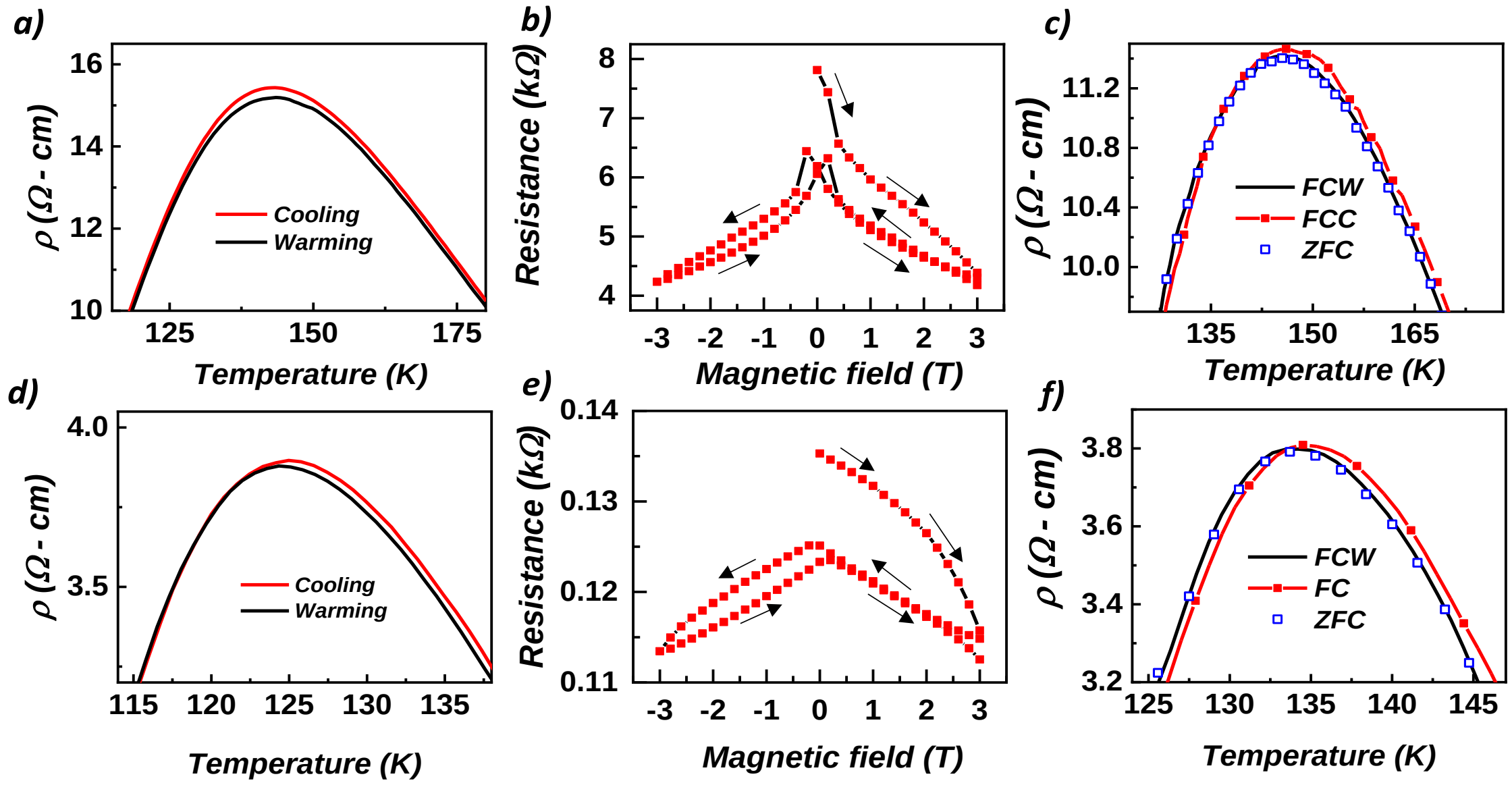

Figure 2 : a), d) Illustrate thermal hysteresis in resistivity during the cooling and warming cycle of NS-G and NS-R b), e) Magnetoresistance plot showing training effect at 4 K in granular and rod-morphology sample. c), f) Illustrate thermal hysteresis in resistivity during the FCC and FCW cycle.

Further, zero-field cooling (ZFC), field-cooled cooling (FCC), and field-cooled warming (FCW) measurements were performed. In the case of ZFC, the samples were cooled in the absence of a magnetic field. Upon reaching 4 K a magnetic field of 0.5 T was applied and the resistivity was recorded during the warming cycle. In the case of FCC and FCW, the resistivity is recorded in the presence of 0.5 T while cooling and warming the sample, respectively. As we take a closer look at the ZFC, FCW, and FCC data, a thermal hysteresis is observed in the FCC and FCW cycle indicating that phase transition is first order in nature and there can be co-existing phases. In a system showing first-order phase transition kinetically arrested phases can exist at temperatures below the transition temperature, and upon warming, these phases are devitrified and it results in a change in the value of physical properties such as resistivity and magnetization.

To further investigate the presence of kinetically arrested phases, we have used the CHUF protocol. To study the kinetically arrested phase in a system with an equilibrium low-temperature phase being FM in nature, the cooling field '$H_{cool}$' is chosen such that $H_{cool} < H_{warming}$ where '$H_{warming}$' is the magnetic field applied during the subsequent warming cycle. This is because cooling in lower magnetic fields gives larger access to the arrested bands in H-T space. For $H_{cool} > H_{warming}$, one can still obtain the arrested phase, but the fraction of the arrested phase is less. With increasing $H_{cool}$, the fraction of the kinetically arrested phase reduces, and the co-existing metastable (supercooled) phase increases. The metastable supercooled phase eventually gets converted to the equilibrium phase with the application of field or lowering the temperature[18–21]. A detailed description regarding the choice of $H_{cool}$ and $H_{warming}$ can be found in the works of Chaddah *et. al.* and Banerjee *et. al*[18–21]. Accordingly, one can tune the fraction of non-equilibrium magnetic phases co-existing with the equilibrium magnetic phase by the application of an external magnetic field using the CHUF protocol[18–21]. Two fundamental features arise when the CHUF protocol is performed in the manganite systems – i) changes in the physical properties due to change in cooling field and ii) a reentrant transition [18–21].

The resistivity and magnetization behavior of the NSMO thin films at low temperatures under the CHUF measurement protocol are shown in Figure 3 and Figure 4. According to the CHUF protocol, the samples were cooled from 300 K to 4 K at different fields (up to 10 T) across the IMT. At 4 K, the field was isothermally switched to a measuring field of 0.5 T. The resistivity/magnetization values were recorded as the sample was warmed to 300 K. The CHUF measurements on NS-G as shown in Figure 3 a) and NS-R in Figure 3 c) reveal that the resistivity decreases significantly and the reduction in resistivity systematically increases upon cooling in a higher magnetic field. The resistance of manganite thin films is expected to decrease with the application of magnetic field as our NSMO thin films show a negative (decreasing) magnetoresistance[16]. This is because the magnetic field tends to align the FM spins which results in the reduced spin-dependent scattering effects, leading to a reduction in resistance. However, even in the magnetoresistance picture, the value of resistance at a given temperature and field does not change much after a complete field cycling (0 T to 5 T to 0 T to -5 T to 0 T). In our case, it is observed that the value of resistivity measured during the 0.5 T warming cycle depends on the thermo-magnetic history, *i.e.*, the value of the magnetic field during the cooling cycle. This provides a visual confirmation of the existence of metastable non-equilibrium phases at low temperatures. The changes in the cooling field have brought about this drastic change in the physical property of resistivity.

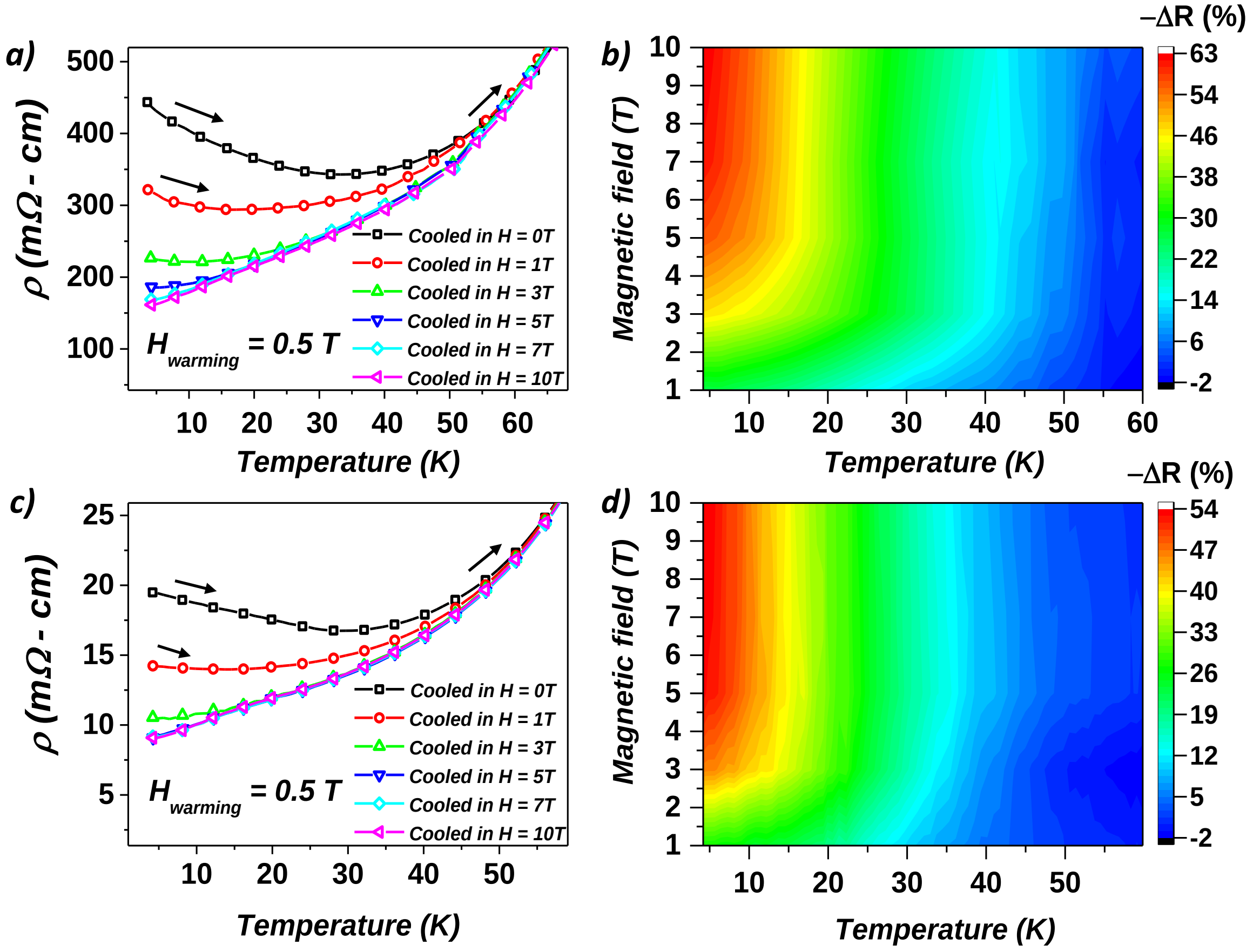

Figure 3 : Resistivity vs. temperature plot of a) NS-G, c) NS-R at low-temperatures (< 60 K ) recorded with warming under 0.5 T magnetic field showing difference in resistivity values as the sample is cooled in different fields. b), d) 3-D plot of estimated difference in resistivity of the 0 T cooled and field cooled case and its evolution with temperature and cooling field for NS-G and NS-R respectively.

Further, a slope change is noted in Figure 3 a), and c), indicated by arrows, for the 0T and 1 T cooled case in both samples. Such a feature is absent and shows a nearly linear fall in resistivity when cooling fields are higher than 1 T. The slope change can be caused by the following factors. In our earlier research, we observed an upturn in manganite thin films as a result of various electron scattering mechanisms at low temperatures[15]; however these effects can be reduced in the presence of magnetic field[32]. Secondly, after cooling the sample in zero field, the FM moments are randomly aligned and subsequent warming in magnetic field causes decreases in resistivity due to polarization of FM moments along the direction of the magnetic field. However, the slope change is evidently observed even in the 1 T cooled case. Therefore the drastic reduction in resistance cannot be accounted for only due to loss of polarization, as the sample warmed in the presence of a magnetic field of 0.5 T. Hence, we corroborate that such a slope change is indicative of devitrification of the kinetic arrested PM phase where a reentrant transition is observed as the sample is warmed. Further, such slope changes are not observed in the case with higher cooling fields. This is because with increasing cooling field, the fraction of arrested PM phase reduces, and the fraction of co-existing supercooled PM phase increases. The supercooled PM phases eventually get transformed to the equilibrium FM phase upon cooling. In consequence, due to reduced spin-dependent scattering effects in the FM phase, the resistivity is drastically reduced at low temperatures (4 K). Upon subsequent warming in 0.5 T, the resistivity increases linearly during the warming cycle. The change in resistivity between the 0 T cooled and the field cooled cases is evaluated as $\Delta R$ (%) and a 3-D graph is plotted in Figure 3 b), d) showing the difference in resistivity with increasing temperature and cooling field for NS-G and NS-R respectively. Although both films show a change in resistivity the maximum percentage drop in resistivity from 0 T cooled case to 10 T cooled is about 64 % in the case of NS-G and 54 % in the case of NS-R. Before we can make remarks on the extent of kinetic arrest across these samples, we examine the change in magnetization values upon CHUF measurements on the NSMO thin films.

Figure 4 a) and b) show the temperature-dependent magnetization curves of NS-G and NS-R measured under the CHUF protocol. Similar to the resistivity curves, it is observed that the magnetization curves diverge below 50 K upon CHUF measurements. Thus, visually confirming the presence of glassy (kinetic arrested and supercooled) PM phases coexisting with the equilibrium FM phase at low temperatures. In contrast to the resistivity case, we observe that the net magnetization increases at low-temperatures as the samples are cooled in higher magnetic fields. Similar to the resistivity case we observe a significant change of slope in the magnetization curves of the 1 T cooled case. As the cooling magnetic field increases, this slope change shifts towards lower temperatures. This shift is due to the decrease in the kinetically arrested phase with the increase in the cooling magnetic field. The difference in magnetization values between the 1 T cooled case and the other field-cooled cases is evaluated as ΔM (%) and the 3D plot showing the variation of ΔM % with temperature and magnetic field is shown in Figure 4 b) and d) for NS-G and NS-R respectively. A maximum ΔM % of only 4 % is obtained in the case of NS-G and ΔM % of 16 % is obtained in the case of NS-R. We explain the enhancement in ΔM % in the case of NS-R with the following reasons. Unlike NS-R, NS-G has a large density of GB defects. Thus, during the phase transition, these defects can act as pinning centres or points of nucleation for the growth of FM phase. As the sample is cooled below the transition temperature, the metastable (supercooled) PM phases get transformed to the equilibrium FM phase readily due to nucleation in NS-G. However, in NS-R, due to reduced GB defects, the metastable PM phase persists up to low temperatures giving rise to a large enhancement in magnetization at 4 K. This indicates unequivocally that the extent of phase coexistence is higher in NS-R as compared to NS-G.

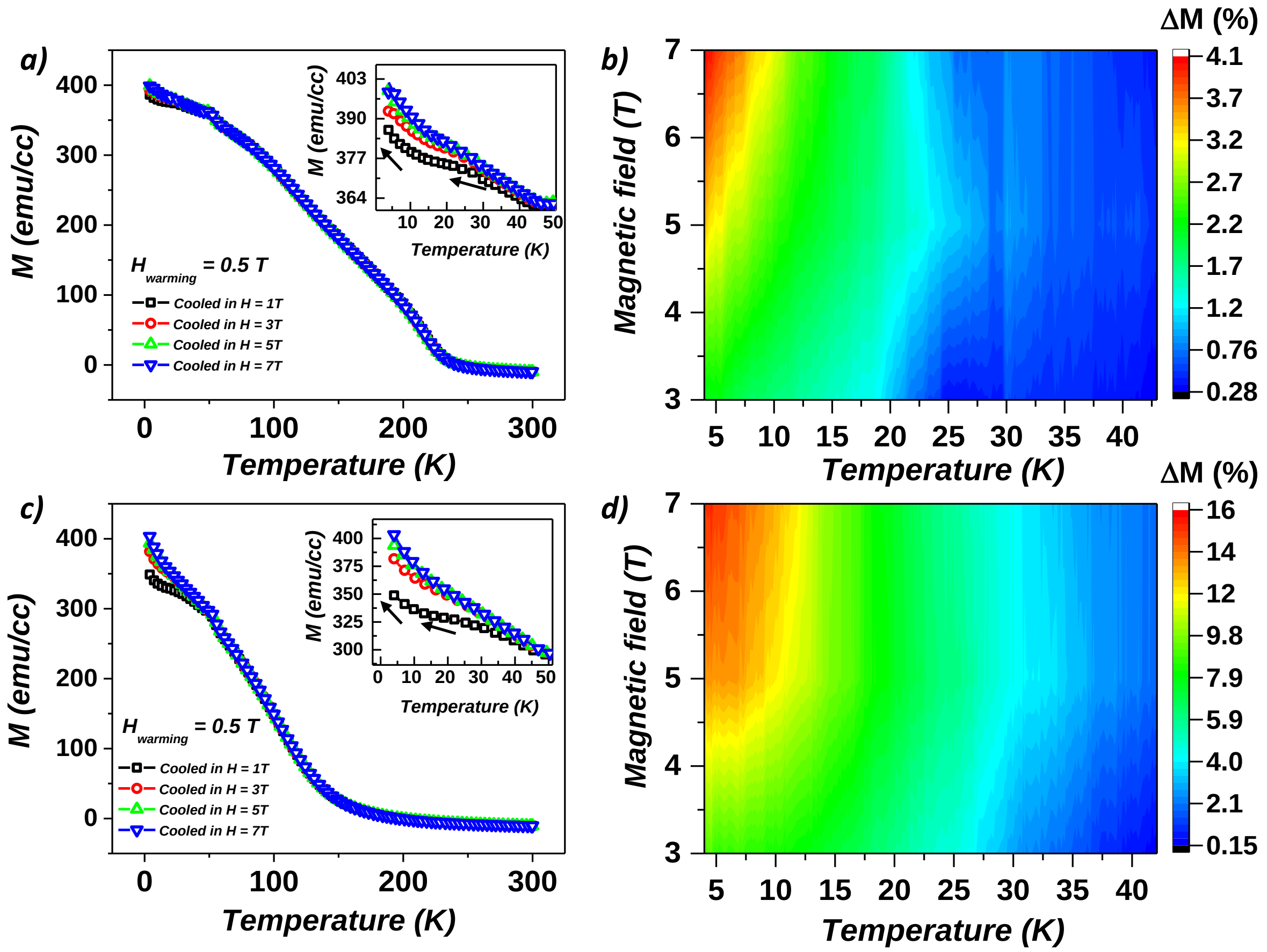

Figure 4 : Magnetization vs. temperature plot of a) NS-G, c) NS-R for a warming field of 0.5 T showing difference in magnetization values as the sample is cooled in different fields. b), d) 3-D plot showing estimated difference in magnetization of the 1 T cooled and field cooled case and its evolution with temperature and cooling field for NS-G and NS-R respectively.

Compared to the resistivity CHUF measurements where NS-G shows a large $\Delta R$ (%), magnetization results indicate the opposite. This may be because the resistivity measurements are sensitive to GB scattering effects thus a large $\Delta R$ (%) is obtained in the case of NS-G compared to NS-R. Further, to gain a deeper understanding of the phase coexistence across these systems, we have performed time-dependent magnetization and magnetotransport studies. This is essential because an additional striking feature of the glassy state is its devitrification. The arrested magnetic phase devitrifies with increasing the temperature at a given magnetic field or upon increasing the magnetic field isothermally [18–21].

Subsequently, the devitrification of glassy magnetic phases towards equilibrium is elucidated through time-dependent measurements using a FC protocol adopted from the works of Chaddah *et. al.* and Tapati Sarkar *et. al.* [33]. Initially the sample is cooled from 300 K to 4 K in a magnetic field of 1 T. The kinetically arrested PM phase coexists with the metastable supercooled PM phase in the glassy phase as the sample is cooled in a 1 T magnetic field. The arrested PM phase remains invariant as the magnetic field is isothermally switched off at 4 K. However, the supercooled non-equilibrium PM state is metastable. Therefore, any energy fluctuation can lead to the conversion of the supercooled PM phase to the equilibrium phase FM in the system[18–21,26]. After a settling time of 300 s the temperature is increased to measuring temperature '$T_m$'. In our case, the measurements were carried out at temperatures starting from 4 K up to 75 K. The magnetization was recorded for 2000 s and the data was fit using the Kohlrausch-Williams-Watt (KWW) [26] stretched exponential function [$M(t) \propto (1 + \exp(-(t/\tau)^{\beta}))$] with $\beta$ values between 0.2 to 0.5 for a relaxation time of the order of $10^4$ s. The normalized magnetization ($M(t)/M(t = 0$ s)) is plotted as a function of time in Figure 5 a), c) for NS-G and NS-R respectively.

At $T_m = 4$ K, the magnetization is observed to decrease with time in both samples. In the given system, the FM moments can lose their polarization and attain the remnant state as the field is switched off at 4 K after cooling in 1 T. However, in such a scenario the magnetization does not change significantly as a function of time. Since the glassy state formed at low temperatures is PM in nature, the decay in magnetization arises from the unsaturated metastable PM phase. Such a trend with decaying magnetization relaxation is observed at each measuring temperature as shown in Figure 5, however, the relaxation rate differs with respect to temperature. The magnetization relaxation shows a non-monotonic behavior as the measuring temperature increases. At $T_m$ = 10 K, the relaxation rate decreases to very low values in both NS-G and NS-R. This decrease in relaxation rate indicates that the supercooled PM phase being metastable is left in the local minima of free energy which prevents it from getting converted to the equilibrium FM phase[33]. The increase in temperature from 4 K to 10 K is believed to increase the barrier height of free energy thereby reducing the relation rate. With further increases in temperature, the nature of relaxation is relatively different across the samples.

In NS-G, the relaxation rate increases with increasing temperature as observed from the 20 K, 30 K, and 40 K curves with the rate being nearly equal to the 4 K case, only with slight variations with increasing temperature. Conversely, in NS-R, the relaxation rate increases systematically for each measuring temperature up to 50 K. The increase in the relaxation rate is attributed to the devitrification of the kinetically arrested phase with increasing temperature. Further, it is confirmed that in both samples the supercooling limit T* is lesser than $T_g$ where $T_g$ is the glass transition temperature. However, the systematic increase in relaxation rate in the NS-R as compared to NS-G hints at the stability of the glassy state across the two systems. As observed from the CHUF results of magnetization, the field-induced formation of a glassy PM state in the FM matrix is different in NS-G as compared to NS-R. Consequently, the devitrification is also different across the samples.

During devitrification of the glassy state, the arrested PM phase gets converted to the equilibrium FM phase, which occurs through nucleation and growth of the FM phase in the matrix of the glassy PM phase. According to Pal. *et.al.*'s work on the devitrification of glassy phase in nano-particles of Pr-Ca-Mn-Al-O whose ground state is FM at low temperatures, the nucleation of equilibrium FM takes place either homogenously or heterogeneously[34]. During homogeneous nucleation, droplets of the FM phase of critical size form randomly within the nanoparticles controlled by the driving energy, and in the heterogeneous case, nucleation of the FM phase takes place around defects such as grain-boundaries[34]. At nanometer length scales these nucleation mechanisms compete with each other[34]. We draw this analogy to our case. The NS-G sample possesses an increased density of GBs and pinning centers as compared to NS-R. Therefore, the nucleation of the FM phase is enhanced around such defects in the case of NS-G. Consequently, the devitrification of the PM phase is evident from that large relaxation rate in NS-G at 20 K. With further increase in temperature, the relaxation rate increases only moderately and this is attributed to the homogenous nucleation mechanism inside the grains in NS-G. However, in NS-R, due to lesser GB defects/pinning centers, the homogenous and heterogeneous nucleation mechanisms compete and the relaxation rate increases steadily with increasing $T_m$ up to 50 K. Around $T_m$ = 50

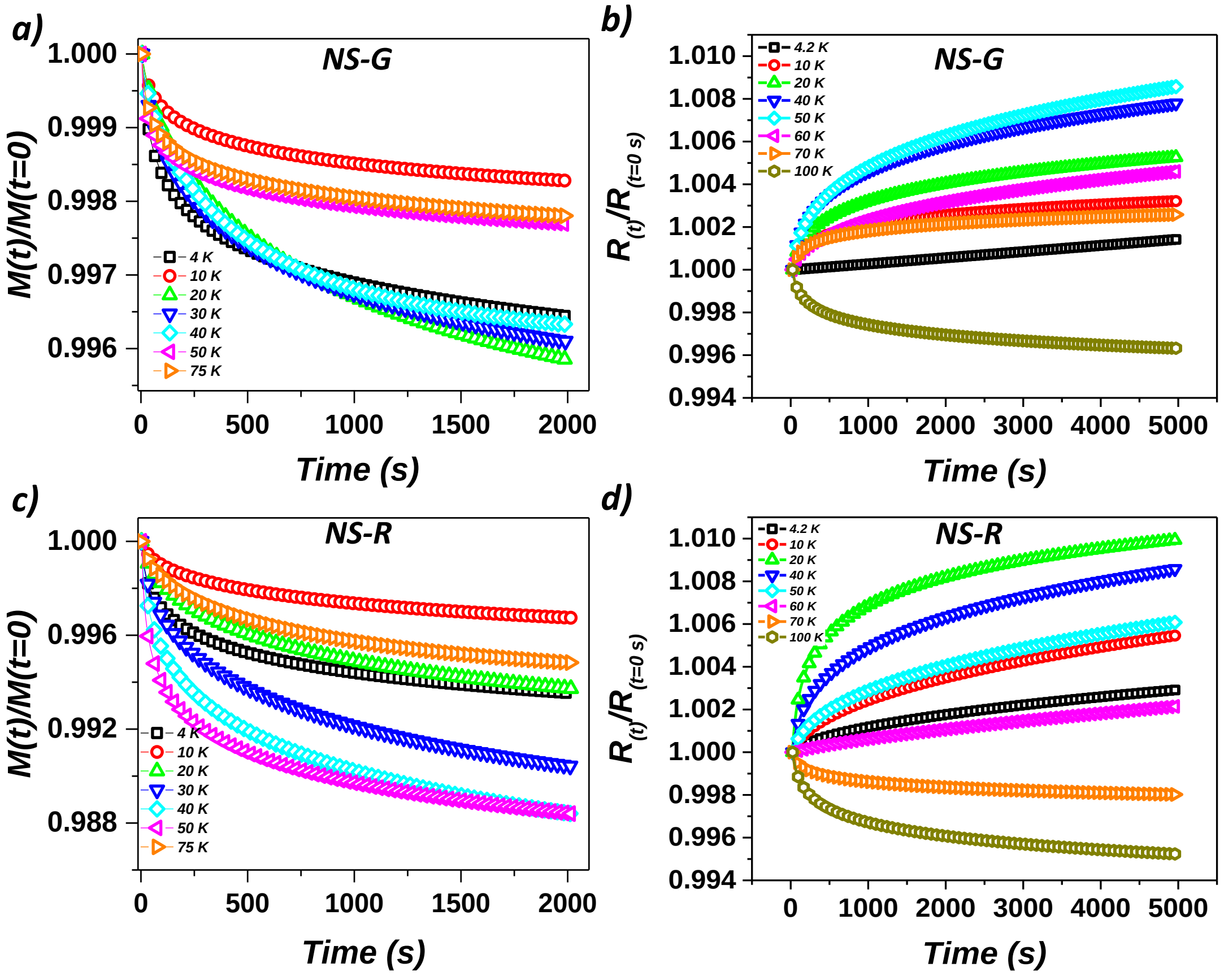

Figure 5 : Time evolution studies using on the NSMO thin films is illustrated. a), b) Magnetization, resistivity measured as a function of time at different measuring temperatures '$T_m$' for NS-G respectively. c), d) Time evolution of magnetization and resistivity at different '$T_m$' for NS-R.

K, and for temperature above 50 K, the relaxation rate drops to very low values. This shows that the glassy PM phase has devitrified completely with increasing temperature ($T_m$ = 75 K) due to faster kinetics. Also, the net change in magnetization in time-dependent measurements is evidently larger in NS-R as compared to NS-G which is in line with CHUF results from magnetization.

The time-dependent FC protocol is repeated during the resistivity measurements for different measuring temperatures and the resistivity was recorded for 5000 s at each measuring temperature. The resistivity data is smoothened and the normalized resistivity (ρ (t)/ρ (t=0)) as a function of time is plotted in Figure 5 b), d). As mentioned earlier, devitrification of the arrested phase cannot take place, as the magnetic field is switched off isothermally at 4 K. Subsequent devitrification can takes place only if the sample is warmed or if the magnetic field increases. Only a marginal increase is observed in resistivity with respect to time at $T_m$ = 4 K for NS-G and NS-R. With further increase in temperature the resistivity time evolution curves for 4 K, 10 K, and 20 K lie systematically one above the other indicating resistivity relaxation rate increases with increasing $T_m$ for both the samples. Corroborating our results from time-dependent magnetization studies, the depolarization of the PM glassy phase as the magnetic field is switched off at 4 K causes an increase in resistivity with time. Further, the enhancement in resistivity relaxation rate with increasing measuring temperature is due to an increased free-energy barrier which causes the supercooled PM phase to remain in the metastable state preventing its conversion to the equilibrium FM phase. Therefore, a net increase in resistivity relaxation rate is observed with time, up to a measuring temperature of 20 K. At $T_m$ = 20 K and above 20 K, the relaxation in resistivity across samples is different. In NS-G, the resistivity relaxation rate increases up to 50 K, whereas in NS-R, the maximum relaxation in resistivity is observed only up to 20 K. This can be explained as follows. With increasing temperature, the kinetically arrested PM phase can devitrify into the equilibrium FM phase resulting in a reduction in the resistivity of the sample. Similar to the case of magnetization, the time-dependent relaxation in resistivity indirectly indicates the stability of the glassy PM phase. However, unlike magnetization which is a bulk measurement, the resistivity of the sample is the result of percolation of conduction of electrons through the FM-metallic phase. Therefore, though the granular sample possesses a large density of pinning centers aiding heterogeneous nucleation of the FM phase, conduction can take place only upon completion of the percolation path. Conversely, the morphology of NS-R aids in the completion of the percolation path during the devitrification of PM phases into the equilibrium FM phase. This results in reduced resistivity and reduction in relaxation rate for $T_m$ > 20 K in NS-R. Whereas, in NS-G the resistivity relaxation rate steadily increases up to $T_m$ = 50 K due to the incomplete percolation path.

Further, as $T_m$ is increased above 50 K it is observed that the direction and nature of relaxation changes. A rapid decrease in the resistivity relaxation curve at $T_m$ > 50 K is observed revealing the complete devitrification of kinetically arrested PM phases. Upon complete devitrification, the system having relaxed into the equilibrium FM state shows a decrease in resistivity with time due to the growth of the low-temperature FM phase at $T_m$ = 75 K. Thus, the magnetization and magnetotransport studies provide a clear picture of the nature of devitrification across the samples with distinct morphologies. The nature of devitrification differs across the samples due to differences in the nucleation process, which arise due to its morphology.

## CONCLUSION

Magnetic and magnetotransport investigations have been carried out using the CHUF protocol to study the magnetic glassy state at low temperatures (< 50 K) in PLD-grown NSMO thin films possessing granular and rod-type morphology. The sample with granular morphology was observed to have a resistivity drop of 64 % and the samples with rod morphology with resistivity drop of 54% upon CHUF measurements. The magnetization measurements using the CHUF protocol revealed that magnetization increases at low temperatures as the samples are cooled in higher magnetic field. The sample with rod morphology shows enhancement in magnetization of about 16 % as compared to the granular sample with a 4 % increase, thus confirming the large extent of phase co-existence in the rod morphology sample. Subsequent measurements involving time-dependent studies on the kinetic arrest were also performed to examine the stability of the magnetic glassy phase across the samples. In comparison to the rod-morphology thin film, the granular NSMO thin film exhibits faster devitrification of the glassy phase during the magnetization measurements, with a maximum relaxation rate at a lower temperature of about 20 K. This is substantiated by the time evolution studies on resistivity. In conclusion, it is observed that the morphology greatly affects the extent of the kinetic arrest and devitrification of the glassy magnetic phase.

## ACKNOWLEDGMENTS

One of the authors (Mrinaleni R S) would like to acknowledge the Department of Atomic Energy, India for providing experimental facilities. We thank UGC-DAE CSR, Kalpakkam node, and for providing access to SQUID magnetometer, magnetotransport measurement system.